\documentclass{article}
\usepackage{graphicx}
\usepackage[english]{babel}
\newcommand{\rec}[1]{\frac{1}{#1}}
\newcommand{\td}[2]{\frac{\mathrm{d}{#1}}{\mathrm{d}{#2}}}
\newcommand{\z}[1]{\left({#1}\right)}
\renewcommand{\ae}[1]{\left|{#1}\right|}
\newcommand{\sz}[1]{\left[{#1}\right]}
\newcommand{\kz}[1]{\left\{{#1}\right\}}
\newcommand{\m}[1]{\mathrm{#1}}

\renewcommand{\v}[1]{\mathbf{#1}}
\renewcommand{\r}[1]{(\ref{#1})}

\title{
\includegraphics[width=0.35\textwidth]{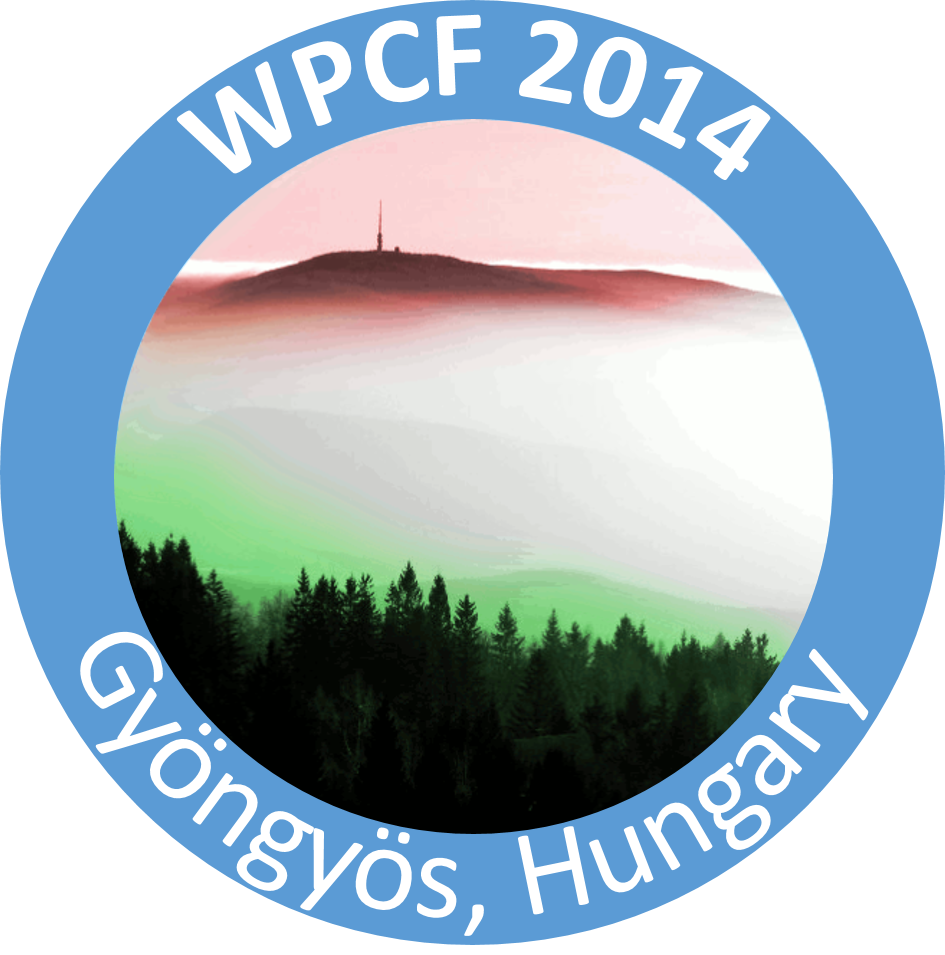}\\[1cm]
An analytic hydrodynamical model of rotating 3D expansion in heavy-ion collisions}
\author{{M.~I.~Nagy$^1$, T. Cs\"org\H o$^{2,3}$,}\\[1ex]
$^1$Dept. of Atomic Physics, ELTE, H-1117 Budapest XI, P\'azm\'any P. 1/A, Hungary\\
$^2$Wigner RCP, H-1525 Budapest 114, POB 49, Hungary\\
$^3$KRF, H-3200 Gy\"ongy\"os, M\'atrai \'ut 36, Hungary
}

\begin{document}
\fontfamily{lmss}\selectfont
\maketitle

\begin{abstract}
A new exact and analytic solution of non-relativistic fireball hydrodynamics is
presented. It describes an expanding triaxial ellipsoid that rotates around one
of its principal axes. The observables are calculated using simple analytic
formulas.  Azimuthal oscillation of the off-diagonal Bertsch-Pratt radii of
Bose-Einstein correlations as well as rapidity dependent directed and third
flow measurements provide means to determine the magnitude of the rotation of
the fireball. Observing this rotation and its dependence on collision energy
may lead to new information on the equation of state of the strongly
interacting quark gluon plasma produced in high energy heavy ion collisions.
\end{abstract}

\section{Introduction}

The applications of hydrodynamics (both numerical and exact analytic solutions) in the description of the ,,soft'' particle
production in heavy-ion collisions has a good tradition. Here we deal with an analytic model. (For a recent work on numerical
hydrodynamics, see eg.\@ Ref.~\cite{Karpenko:2013wva}; the many references therein provide a good list of recent developments
in numerical hydrodynamical modelling). Exact solutions require some ,,luck'' to find, however, they also have a good line of
success: the early days of high energy physics saw the development of the Landau-Khalatnikov solution~\cite{Landau} and the
Hwa-Bjorken solution~\cite{HwaBjorken}, both invaluably useful for high energy collision phenomenology. Since then, many
physically realistic solutions have been found, relativistic as well as non-relativistic ones.

The rotation of the expanding hot and dense matter produced in non-central heavy ion collisions drew much theoretical attention
recently. Many numerical models try to grasp the effects of rotation on the observables (see eg.
Refs.~\cite{Csernai:2013vda,Csernai:2014rva}). In this work we present an exact solution of the non-relativistic perfect fluid
hydrodynamical equations that describes the desired rotating expansion of a triaxial ellipsoid-like fireball, just as one
imagines the space-time picture of a non-central heavy-ion collision. We also calculate the observables using simple analytic
formulas and point out those features of them that would reveal the magnitude of rotation.

We mention some of our works of which this step is a natural continuation: the discovery of accelerating relativistic
solutions~\cite{Csorgo:2006ax,Nagy:2007xn} as well as the first rotating relativistic~\cite{Nagy:2009eq} and
non-relativistic~\cite{Csorgo:2013ksa} exact solutions. There are also exact solutions available for arbitrary equation
of state in the relativistic~\cite{Csanad:2012hr} domain (even ones containing multipole asymmetries~\cite{Csanad:2014dpa})
as well as in the non-relativistic domain~\cite{Csorgo:2001xm} with realistic, ellipsoid-like symmetry. The solution and the
calculation of the observables presented below is thus a next example of usefulness of exact solutions in heavy-ion phenomenology.
The current work is a straightforward, although rather non-trivial generalization of our recent work on the evaluation of the observables
for a spheroidally symmetric, rotating and expanding family of exact solutions of fireball hydrodynamics 
~\cite{Csorgo:2015scx} for more general expansions with triaxial ellipsoidal symmetry. We refer to this work also  for a more detailed
overview and introduction to the status of the field of exact solutions of fireball hydrodynamics~\cite{Csorgo:2015scx}.

\section{A new rotating solution of hydrodynamics}

The equations of non-relativistic hydrodynamics are the Euler equation, the energy conservation equation and the particle number
conservation equation. They are recited here in the form suitable for heavy-ion physics phenomenology:
\begin{eqnarray}
nm_0\z{\partial_t +\v v\nabla}\v v &=& -\nabla p, \label{e:euler} \\
\z{\partial_t +\v v\nabla}\varepsilon &=& -\z{\varepsilon + p}\nabla\v v, \label{e:energy} \\
\z{\partial_t +\v v\nabla} n &=& - n\nabla\v v. \label{e:ncont}
\end{eqnarray}
Here $n$ stands for the particle number density (thus $nm_0$ is the mass density), $T$ for temperature, $p$ for pressure, $\v v$ for
the velocity field, $\varepsilon$ for the energy density. The equations above need to be supplemented with a suitable Equation of
State (EoS) which we customarily choose as
\begin{eqnarray}
\varepsilon &=& \kappa\z{T}p,\\
p &=& nT.
\end{eqnarray}
In non-central heavy-ion collisions, an almond-shaped region of hot and dense matter forms which has non-zero angular momentum. The
almond shape is approximated with a triaxial ellipsoid, and the following solution of hydrodynamics takes rotation and expansion
into account. Let the $x$ axis point in the direction of the impact parameter and the $z$ axis in the direction of the colliding
beams; thus the $x$--$z$ plane is the event plane, and the rotation is around the $y$ axis. We call the lab frame the $K$ frame,
to be distinguished from the $K'$ frame, which co-rotates with the principal axes of the expanding ellipsoid. The solution is
described in terms of the time-dependent angle of rotation $\vartheta\z{t}$, and the three principal axes of the rotated
ellipsoid $X\z{t}$, $Y\z{t}$, $Z\z{t}$. The scaling variable, whose level surfaces correspond to self-similar ellipsoids
is\footnote{In the following, we do not explicitly write up the time dependence of functions which were introduced as
time-dependent ones.}
\begin{equation}\label{e:sdef}
s\z{\v r,t} \equiv
\frac{{r'}^2_x}{X^2}+
\frac{{r'}^2_y}{Y^2}+
\frac{{r'}^2_z}{Z^2} ,
\end{equation}
\begin{equation}\label{e:KK1transf}
\begin{array}{rcl} 
r'_x &=& r_x\cos\vartheta - r_z\sin\vartheta,\\
r'_y &=& r_y,\\
r'_z &=& r_x\sin\vartheta + r_z\sin\vartheta.
\end{array}
\end{equation}
The velocity field is taken to be a linear, rotating Hubble-like one, for which the above $s$ variable is a proper ,,scaling
variable'', ie.\@ its co-moving derivative vanishes. We introduce the $V = XYZ$ notation. One can directly check that for
constant $\kappa$ the above and the following formulas provide the desired rotating solution to the \r{e:euler}--\r{e:ncont}
hydrodynamical equations:
\begin{equation}\label{e:vdef}
\v v'\z{\v r,t} = \z{\begin{array}{l}
\frac{\dot X}{X}r'_x + g\frac{X}{Z}r'_z\vspace{1mm} \\ \frac{\dot Y}{Y}r'_y\vspace{1mm}  \\ \frac{\dot Z}{Z}r'_z - g\frac{Z}{X}r'_z \end{array}} ,
\end{equation}
\begin{equation}\label{e:dotthetadef}
\dot\vartheta\z t = g\z t = \frac{\omega_0}{2}\z{\frac{X_0+Z_0}{X+Z}}^2,
\end{equation}
\begin{eqnarray}
n &=& n_0\frac{V_0}{V}{\mathcal{V}}\z s,\label{e:ndef} \\ 
T &=& T_0\z{\frac{V_0}{V}}^{\rec\kappa}{\mathcal{T}}\z s,\label{e:Tdef} \\ 
{\mathcal{V}}\z{s} &=& \rec{{\mathcal{T}}\z{s}}e^{-\rec 2\int_0^s\frac{\m ds'}{{\mathcal{T}}\z{s'}}} .\label{e:nutau} \\ 
\end{eqnarray}
For non-constant $\kappa$, the above formulas can be modified to form a valid solution. A solution is found with spatially
homogeneous $T$ and Gaussian-like $n$ as
\begin{equation}\label{e:nsol}
n = n_0\frac{V_0}{V}e^{-s/2},
\end{equation}
\begin{equation}\label{e:kappaTsol}
T\equiv T\z t,
\quad\frac{V_0}{V}=\exp\kz{\int_{T_0}^T\m d\beta\z{\td{\kappa}{\beta}+\frac{\kappa}{\beta}}} .
\end{equation}
Furthermore, the time development of the $X$, $Y$, $Z$ principal axes (and thus that of $\dot\vartheta$ and $g$ is governed
by a Lagrangian for a point-like particle of mass $m_0$, which we write up only in the $\kappa = const$ , $ T\equiv T\z t$ case:
\begin{equation}\label{e:L}
L = \frac{m_0}{2}\z{\dot X^2+\dot Y^2+\dot Z^2} - \rec\kappa\z{\frac{T}{T_0}} - 
m_0 \omega^2 R_0^2,
\end{equation}
where we introduced another simplifying notation: the ,,average'' radius of the ellipsoid $R$, and the ,,average'' angular velocity $\omega\z t$ as
\begin{equation}
R\equiv \frac{X+Z}{2},\quad 
R_0\equiv \frac{X_0+Z_0}{2},\quad 
\omega\z{t} \equiv \omega_0\frac{R_0^2}{R^2} .
\end{equation}
Note that the velocity profile is written up in the co-rotating $K'$ frame. The free parameter $\omega_0$ is related to the
total angular momentum $M_y$ of the flow (which points in the $y$ direction); for example, in the ${\mathcal{V}}\z{s} = e^{-s/2}$ Gaussian case
\begin{eqnarray}\label{e:My}
M_y & = & 2 M_0 R_0^2 \omega_0 , \\
M_0 & = & m_0 \z{2\pi}^{3/2} V_0 n_0 .
\end{eqnarray}
The solution thus specified is a natural generalization of the earlier results for  three dimensionally expanding, non-rotating
triaxial ellipsoids~\cite{Csorgo:2001xm,Csorgo:2001ru} (indeed, our formulation resembles very much to those ones),  and in the spheroidal limit
$(X = Z) $ it reproduces the exact solutions presented in Refs.~\cite{Csorgo:2013ksa,Csorgo:2015scx} for rotating and exploding fireballs with spheroidal symmetry.

\section{Observables from the new solution}

One can calculate the hadronic observables from the above solution by specifying the emission function (source function) and a
suitable freeze-out condition. Following Ref.~\cite{Csorgo:2001xm}, for simplicity (and because the solution for arbitrary
$\kappa\z T$ is then available), we take a Gaussian ansatz for the density profile (thus the temperature is spatially
homogeneous), as well as take the freeze-out to happen at a constant time $t_f$ (which in this case corresponds to a
given $T_f$ freeze-out temperature). We assume that at the freeze-out, particles with mass $m$ are created. The source
function is then taken to be the non-relativistic Boltzmann distribution for a particle with mass $m$:
\begin{equation}\label{e:Sdef}
S\z{\v r,\v p} = \frac{n\z{t_f,\v r}}{\z{2\pi mT_f}^{3/2}}\exp\kz{-\frac{\z{\v p-m\v v\z{t_f, \v r}}^2}{2mT\z{t_f}}} .
\end{equation}
In the following, the $_f$ index means the value taken at the freeze-out time $t_f$, but it is mostly dropped: all quantities
are to be understood as their value at $t_f$. The single-particle spectrum and the two-particle Bose-Einstein correlation
function are calculated as
\begin{equation}\label{e:N1def}
N_1\z{\v p} = \int\m d^3\v r\,S\z{\v r,\v p},
\end{equation}
\begin{equation}\label{e:C2def}
C\z{\v K,\v q} = 1 + \lambda\frac{\ae{\tilde S\z{\v K,\v q}}^2}{\ae{\tilde S\z{\v K,\v 0}}^2} ,
\end{equation}
where $\v K$ and $\v q$ are the average pair momentum and the relative momentum, respectively, and $\lambda$ is the effective
intercept parameter of the correlation function. For the mentioned Gaussian density case, all the integrals can be performed
analytically, yielding simple results. We introduce the following quantities for convenience:
\begin{eqnarray}
T^*_{xx} 
= T_f + m (\dot X^2 +  \omega^2 R^2) , \\
T^*_{zz} 
= T_f + m (\dot Z^2 +  \omega^2 R^2), \\
T^*_{xz} 
 = m \omega R\z{\dot X-\dot Z} .
\end{eqnarray}
Note that these definitions can be straightforwardly specialized to the case when the two principal axes $X$ and $Z$ are equal
($X = Z = R$), and thus one deals with the expansion of a rotating spheroid.

The single particle spectrum can be then written as
\[
\td{n}{^3\v p}\propto \exp\z{-\frac{p_x^2}{2mT_x}-\frac{p_y^2}{2mT_y}-\frac{p_z^2}{2mT_z}-\frac{\beta_{xz}}{m}p_xp_z} =
\]
\begin{equation}\label{e:N1}
= \exp\z{-\frac{{p'}_x^2}{2mT'_x}-\frac{{p'}_y^2}{2mT'_y}-\frac{{p'}_z^2}{2mT'_z}-\frac{\beta'_{xz}}{m}p'_xp'_z},
\end{equation}
where $T_x$, $T_y$, $T_z$, and $\beta_{xz}$, as well as $T'_x$, $T'_y$, $T'_z$ and $\beta'_{xz}$ characterize the slope parameters
(and the cross-terms) in the $K$ (laboratory) frame and the $K'$ frame (the eigen-frame of the rotated ellipsoid), respectively.
The expression of these parameters are given as
\begin{eqnarray}
T'_x        &=& T^*_{xx} - \frac{{T^*}^2_{xz}}{T^*_{zz}},\\
T'_z        &=& T^*_{zz} - \frac{{T^*}^2_{xz}}{T^*_{xx}},\\
\beta'_{xz} &=& \frac{T^*_{xz}}{T^*_{xx}T^*_{zz}-{T^*}^2_{xz}},
\end{eqnarray}
\begin{eqnarray}
\rec{T_x}  &=& \frac{\cos^2\vartheta}{T'_x}+\frac{\sin^2\vartheta}{T'_z}+\beta'_{xz}\sin\z{2\vartheta} ,\\
\rec{T_z}  &=& \frac{\sin^2\vartheta}{T'_x}+\frac{\cos^2\vartheta}{T'_z}-\beta'_{xz}\sin\z{2\vartheta} ,\\
\beta_{xz} &=& \frac{\sin\z{2\vartheta}}{2}\z{\rec{T'_z}-\rec{T'_x}}+\beta'_{xz}\cos\z{2\vartheta} ,
\end{eqnarray}
\begin{equation}\label{e:Ty}
T'_y = T_{y} = T + m\dot Y^2 .
\end{equation}
We see that the formulas relating the parameters in the $K$ and $K'$ frames are simply the ones describing a rotation by a
fixed $\theta$ angle (the value of the tilt angle at freeze-out). The fact that a cross term (ie. nonzero $\beta'_{xz}$)
appears (this is a new feature compared to Ref.~\cite{Csorgo:2001xm}) signals that the single-particle spectrum is \emph{not}
diagonal in the $K'$ frame: the tilt of the coordinate-space ellipsoid is not the same as that of the ellipsoid determined from
the single-particle spectrum.

\begin{figure}[ht]%
\begin{center}
\includegraphics[width=0.7\columnwidth]{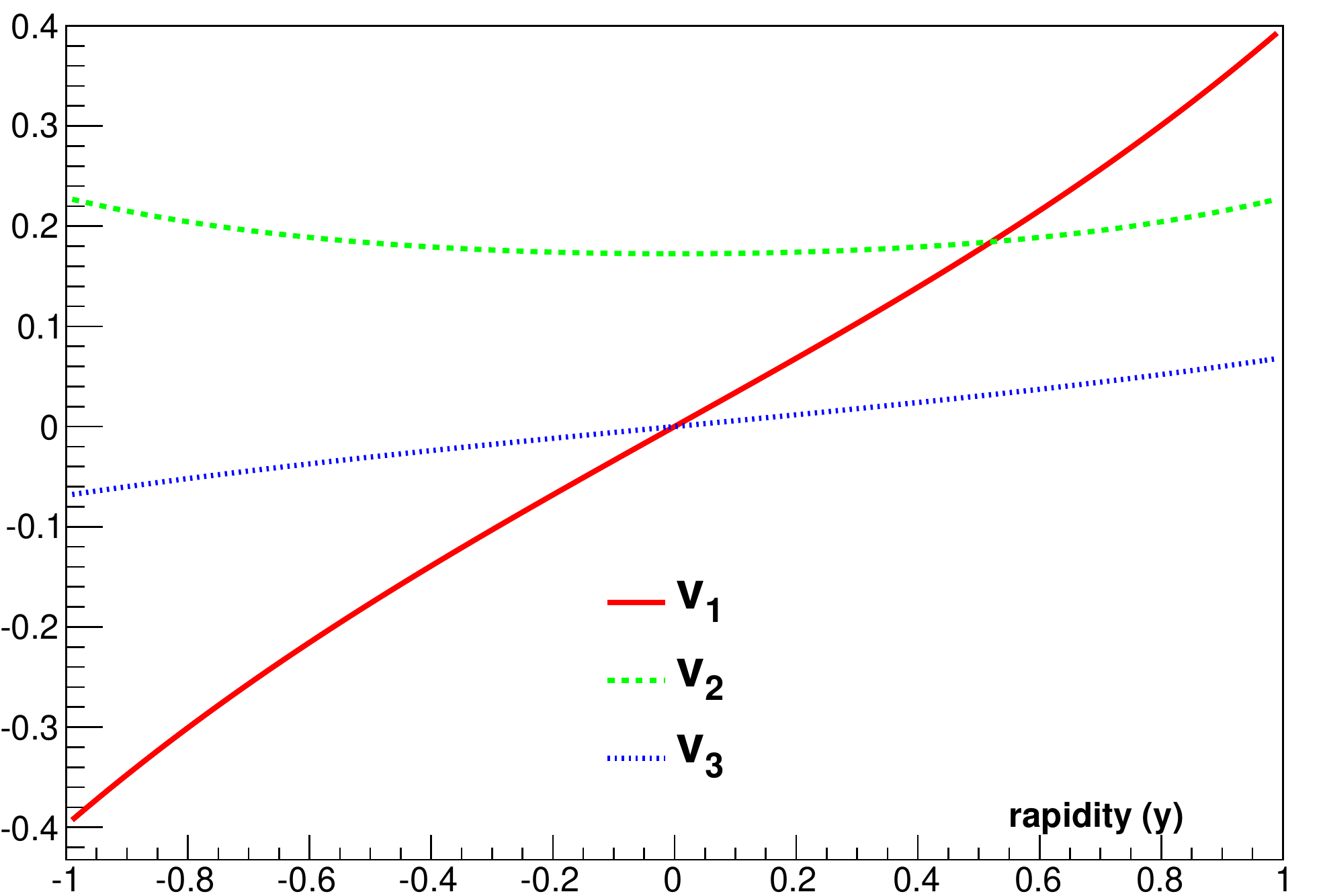}%
\caption{Dependence of the azimuthal anisotropy parameters $v_1$, $v_2$ and $v_3$ on rapidity $y$, for a reasonable choice of the
freeze-out parameters (namely: $\vartheta_f = \pi/5$, $\beta'_{xz}=5\cdot 10^{-4}$ 1/MeV, $p_T = 300$ MeV$/c$, $m=140$ MeV,
$T_x = 270$ MeV, $T_y = 170$ MeV, $T_z = 600$ MeV). The dependence of $v_1$ and $v_3$ is characteristic to a system tilted
by some $\vartheta_f$ angle.}%
\label{f:vn}%
\end{center}
\end{figure}

\begin{figure}%
\begin{center}
\includegraphics[width=0.7\columnwidth]{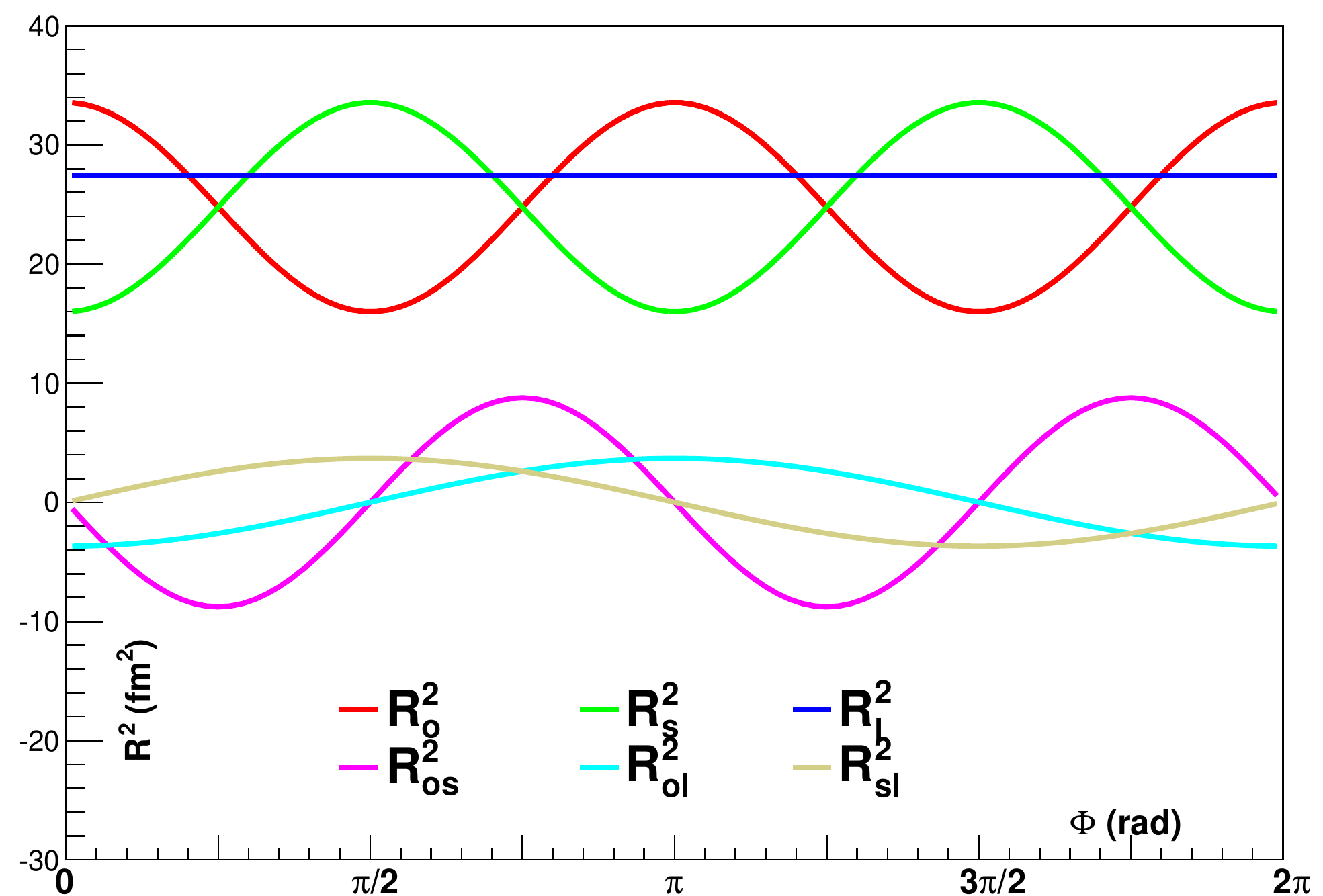}%
\caption{$\varphi$ dependence of the Bertsch-Pratt radii for a reasonable choice of the freeze-out parameters (namely:
$\vartheta_f = \pi/5$, ${R'}^2_x = 25$ fm$^2$, ${R'}_y^2 = 16$ fm$^2$, ${R'}_z^2 = 36$ fm$^2$, ${R'}_{xz}^2 = 5$ fm$^2$).
The oscillation of $R_s$, $R_o$ and $R_{os}$ with a $\pi$ period is characteristic to an ellipsoid-shaped source.
The $2\pi$ period oscillation of $R_{ol}$ and $R_{sl}$ are characteristic to a tilted (and rotating) source.}%
\label{f:HBT}%
\end{center}
\end{figure}

The expression of the azimuthal angle-averaged $p_T$ spectrum and the anisotropy ($v_n$) parameters follow the footsteps of
Ref.~\cite{Csorgo:2001xm}: one can introduce the scaling variables $v$ and $w$ and the effective temperature $T_{\m{eff}}$ as
\begin{equation}\label{e:wnuTeff}
v\equiv-\frac{\beta_{xz}}{m}p_zp_T,\quad
w\equiv \frac{p_T^2}{4m}\z{\rec{T_y}-\rec{T_x}},\quad
\rec{T_{\m{eff}}} = \rec 2\z{\rec{T_y} + \rec{T_x}} .
\end{equation}
Using the $I_n\equiv I_n\z{w}=\rec\pi\int_0^\pi\cos\z{n\varphi}e^{w\cos\varphi}\m d\varphi$ modified Bessel functions of $w$, up
to second order in $v$ (and thus in the $y$ rapidity) one has 
\begin{eqnarray}\label{e:pTvn}
E\td{n}{^3\v p} & =&  \rec{2\pi p_T}\td{n}{p_T\m dy}\kz{1+2\sum_{n=1}^\infty v_n\cos\sz{n\z{\varphi-\Psi_n}}}\quad\Rightarrow , \\
\rec{2\pi p_T}\td{n}{p_T\m dy}&\propto & \exp\kz{-\frac{p_z^2}{2mT_z}-\frac{p_T^2}{2mT_{\m{eff}}}}\times\sz{I_0\z w + \frac{I_0\z w+I_1\z w}{4}v^2}, \\
v_1 &=& \frac{v}{2}\frac{I_0\z w+I_1\z w}{I_0\z w},\\
v_2 &=& \frac{I_1\z w}{I_0\z w}+\frac{v^2}{8}\sz{1+\frac{I_2\z w}{I_0\z w}-2\frac{I_1^2\z w}{I_0^2\z w}},\\
v_3 &=& \frac{v}{2}\frac{I_1\z w+I_2\z w}{I_0\z w}.
\end{eqnarray}
Here $\varphi$ denotes the azimuthal angle, $\Psi_n$ is the $n$th order event plane (in our model, all the event planes coincide,
ie.\@ fluctuation effects are not taken into account). Fig.~\ref{f:vn} presents the characteristic rapidity dependence of $v_1$,
$v_2$ and $v_3$.
The two-particle Bose-Einstein correlation also turns out to be a Gaussian, and can either be expressed in terms of the $\v q'$ of
the $\v q$ relative momentum (in the $K$ and $K'$ frames, respectively), or using the Bertsch-Pratt parametrization. Simple
calculation yields
\begin{eqnarray}\label{e:Cq}
C\z{\v K,\v q} & = & 1 +\lambda\exp\kz{-\sum_{ij}q'_iq'_j{R'}^2_{ij}} \, = \, 1 +\lambda\exp\kz{-\sum_{ij}q_iq_jR^2_{ij}} , \\
{R'}^2_x    &=& X^2\frac{T_f}{T'_x} , \\
{R'}^2_z    &=& Z^2\frac{T_f}{T'_z} , \\
{R'}^2_{xz} &=& XZT_f\beta'_{xz} ,
\end{eqnarray}
where we utilized the already calculated values of the $T'_x$, $T'_z$, $\beta'_{xz}$ slope parameters. These radii are valid in the
$K'$ frame; transformation into the $K$ frame (lab frame) is straightforward:
\begin{eqnarray}
R^2_x    &=& {R'}^2_x\cos^2\vartheta + {R'}^2_z\sin^2\vartheta + {R'}^2_{xz}\sin\z{2\vartheta}, \\
R^2_z    &=& {R'}^2_x\sin^2\vartheta + {R'}^2_z\cos^2\vartheta - {R'}^2_{xz}\sin\z{2\vartheta}, \\
R^2_{xz} &=& \frac{\sin\z{2\vartheta}}{2}\z{{R'}^2_x-{R'}^2_z}+\cos\z{2\vartheta}{R'}^2_{xz}.
\end{eqnarray}
The $y$ direction, as before, is much simpler:
\begin{equation}\label{e:Ry}
{R'}^2_y   = R^2_y = Y^2 \frac{T_f}{T_y} .
\end{equation}
Customarily introducing the $o$, $s$, $l$ (out, side, long) Bertsch-Pratt components of $\v q$ in the longitudinally co-moving system
(LCMS), one has
\begin{equation}\label{e:CqBP}
C\z{\v K,\v q} = 1 +\lambda\exp\kz{-\sum_{a,b=o,s,l}q_aq_bR^2_{ab}} ,
\end{equation}
and the radii as functions of the pair momentum azimuthal angle $\varphi$ are
\begin{equation}\label{e:RBP}
\begin{array}{rl}	
{R'}^2_{oo} \equiv& R^2_o = R_x^2\cos^2\varphi + R_y^2\sin^2\varphi,\\
{R'}^2_{ss} \equiv& R^2_s = R_x^2\sin^2\varphi + R_y^2\cos^2\varphi,\\
{R'}^2_{ll} \equiv& R^2_l = R_z^2,\\
{R'}^2_{os} = & \z{R_y^2-R_x^2}\sin\varphi\cos\varphi,\\
{R'}^2_{ol} = & R^2_{xz}\cos\varphi,\\
{R'}^2_{sl} = & R^2_{xz}\sin\varphi.
\end{array}
\end{equation}
Fig.~\ref{f:HBT} shows the Bertsch-Pratt radii as a function of pair momentum azimuthal angle. The 2nd order oscillating
($\cos^2\varphi$, $\sin^2\varphi$ containing) BP-radii, also the cross-terms, were recently extensively measured by the
STAR collaboration~\cite{Adamczyk:2014mxp}, however, as we see, it is not these oscillations that signal rotational expansion.

\section{Discussion and outlook}

Most of the formulae of the observables derived here show a very close similarity (even identity) to those calculated in
Ref.~\cite{Csorgo:2001xm}, where a \emph{stationary, but tilted} ellipsoidal source was assumed. The universal scaling
of the elliptic flow and other anisotropies (a feature of experimental data that was successfully explained in terms
of the Buda-Lund model~\cite{Csanad:2008af}) is preserved here, just the scaling variables are expressed in a bit more
involved way, see Eq.~\r{e:wnuTeff}. The oscillation of the HBT radii characteristic to rotation also shows up in other
models of tilted ellipsoidal sources (just as in Ref.~\cite{Csorgo:2001xm}; a simpler model was introduced earlier in
Ref.~\cite{Lisa:2000ip}).

The new result in the present work is two-fold: first, an actual (and in some sense, unique) rotating and expanding
hydrodynamical solution is found which naturally leads to tilted sources from a non-tilted initial condition (and
it can be applied to follow the time-evolution of rotation, for any given $\kappa\z T$ EoS). Second, the result for
the observables are refined in a way that takes not only the tilt but also the rotational flow into account.

The most striking consequence of tilted (and also rotating) expansion are seen in the rapidity dependence of the
anisotropy parameters (Fig.~\ref{f:vn}) as well as the \emph{directed flow-like} (ie.\@ azimuthally $2\pi$-periodic)
oscillations of the $R_{sl}$ and $R_{ol}$ HBT radius parameters (the appearance of these cross-terms is purely due
to rotation, see Eq.~\r{e:RBP}). A natural proposal is thus to measure these observables to infer the rotational
angle $\vartheta_f$ and possibly the angular velocity. It is worthwhile to mention that to do these measurements,
one needs experimental information on not only the usual (2nd order) reaction plane but also on the first order
reaction plane; this may be done precisely by utilizing the $y$-dependence of the $v_1$ anisotropy parameter. It
is also worth to mention that in the case of models with spheroidal expansion (that was discussed eg.~in
Refs.~\cite{Csernai:2013vda,Csorgo:2013ksa}), the ,,angle of rotation'' $\vartheta_f$ becomes ill-defined,
so the experimental signatures of rotation become much harder to be identified. 

The detection of rotation has some far-going promises. A softer equation of state (caused eg.\@ by the presence
of a critical endpoint on the QCD phase diagram) would mean that the matter expands less violently, thus the angle
of rotation will be greater. In this manner, measuring the rotation angle as a function of collision energy can be
of great importance in mapping out the critical endpoint and the location of quark-hadron phase transition. So we
look forward to see whether our presented model is applicable to new collision-energy dependent measurements of
HBT radii oscillations as well as rapidity dependent anisotropy parameters to infer the rotation of the expanding
hot and dense matter. It would lead to new knowledge of the strongly interacting quark-gluon-plasma produced in
heavy-ion collisions.

This work was supported by the Hungarian OTKA grant NK101438. M.~N. was supported by the T\'AMOP 4.2.4. A/1-11-1-2012-0001
,,National Excellence Program'', financed by the European Union and the State of Hungary, co-financed by the European
Social Fund.

\end{document}